\documentclass[a4paper,fleqn]{article}
\usepackage{amsmath}
\begin{document}

\title{\textbf{On a modified Vakhnenko--Parkes equation}}

\author{\textsc{Sergei Sakovich}\bigskip \\
\small Institute of Physics, National Academy of Sciences of Belarus \\
\small sergsako@gmail.com}

\date{}

\maketitle

\begin{abstract}
We show that the modified Vakhnenko--Parkes equation, introduced recently by Wazwaz, is an avatar of the sine-Gordon equation.
\end{abstract}

\section{Introduction}

Recently, Wazwaz \cite{W} introduced the modified Vakhnenko--Parkes (MVP) equation
\begin{equation}
u u_{xxt} -u_x u_{xt} + u^3 u_t = 0 , \label{e1}
\end{equation}
similar in its form to the original Vakhnenko--Parkes equation \cite{VP}
\begin{equation}
u u_{xxt} -u_x u_{xt} + u^2 u_t = 0 . \label{e2}
\end{equation}
Wazwaz \cite{W} showed that the MVP equation \eqref{e1} passes the Painlev\'{e} test for integrability, in the formulation for PDEs \cite{WTC,T}, and possesses a three-soliton solution. The reliability of the Painlev\'{e} test has been empirically verified by analysis of numerous multi-parameter families of nonlinear equations (see, e.g., \cite{HO}--\cite{X}, to mention a few), while the existence of a three-soliton solution is also generally considered as a clear indicator of integrability. Therefore it interesting to investigate whether---and in which sense---the MVP equation \eqref{e1} is integrable.

\section{Integrability} \label{s2}

The MVP equation \eqref{e1} is a third-order two-dimensional PDE, therefore its general solution must contain three arbitrary functions depending on one variable each. One of those arbitrary functions is evidently related to the invariance of the MVP equation \eqref{e1} with respect to the transformation $t \mapsto f(t)$ with any function $f(t)$. In other words, if a function $u(x,t)$ satisfies \eqref{e1}, then $u(x,f(t))$ will do for any $f(t)$. This kind of invariance can be useful to lower the order of a studied equation \cite{SS}--\cite{S6}.

Let us multiply \eqref{e1} by $2 u_{xt} / u^3$ and integrate with respect to $x$. Then we get
\begin{equation}
\frac{u_{xt}^2}{u^2} + u_t^2 = g(t) , \label{e3}
\end{equation}
where $g(t)$ is an arbitrary function.

If $g(t) \ne 0$ in \eqref{e3}, we can make $g(t) = 1$ by the transformation $t \mapsto f(t)$ with an appropriately chosen function $f(t)$ (of course, we consider complex-valued transformations and variables throughout). Then we get \eqref{e3} in the form
\begin{equation}
u_{xt} = u \left( 1 - u_t^2 \right)^{1/2} , \label{e4}
\end{equation}
where we have omitted the $\pm$ sign in the right-hand side because the sign can be changed by $t \mapsto - t$. This equation \eqref{e4} is related to the sine-Gordon equation
\begin{equation}
v_{xt} = \sin v \label{e5}
\end{equation}
by the potential transformation
\begin{equation}
u = v_x (x,t) . \label{e6}
\end{equation}
Since the function $g(t)$ in \eqref{e3} is arbitrary, the corresponding function $f(t)$ in the used transformation  $t \mapsto f(t)$ is arbitrary (but non-constant) as well. Consequently, all solutions of the MVP equation \eqref{e1} which correspond to non-zero $g(t)$ in \eqref{e3} are given by the expression
\begin{equation}
u = v_x (x,h(t)) , \label{e7}
\end{equation}
where $h(t)$ is an arbitrary non-constant function (inverse to $f$), and $v(x,t)$ stands for any solution of the sine-Gordon equation \eqref{e5}.

If $g(t) = 0$ in \eqref{e3}, we have
\begin{equation}
u_{xt} \pm i u u_t = 0 , \label{e8}
\end{equation}
where $i^2 = -1$. The potential transformation
\begin{equation}
u = \pm i w_x (x,t) \label{e9}
\end{equation}
relates this equation \eqref{e8} to the Liouville equation
\begin{equation}
w_{xt} = \exp w , \label{e10}
\end{equation}
whose general solution is well known:
\begin{equation}
w = \log \frac{2 a_x b_t}{(a+b)^2} , \label{e11}
\end{equation}
where $a(x)$ and $b(t)$ are arbitrary non-constant functions. Consequently, all solutions of the MVP equation \eqref{e1} which correspond to $g(t) = 0$ in \eqref{e3} are given by the expression
\begin{equation}
u = \pm i \left( \frac{a_{xx}}{a_x} - \frac{2 a_x}{a+b} \right) , \label{e12}
\end{equation}
where $a(x)$ and $b(t)$ are arbitrary non-constant functions.

At last, when we used the integrating factor $2 u_{xt} / u^3$ to obtain \eqref{e3} from \eqref{e1}, the factor was assumed to be non-zero. If $u_{xt} = 0$, we have $u_t = 0$ from \eqref{e1}, that is
\begin{equation}
u = c(x) , \label{e13}
\end{equation}
where $c(x)$ is an arbitrary function. This class of solutions of \eqref{e1} is (formally) covered by the class \eqref{e12} if we set the function $b(t)$ to be a constant.

\section{Conclusion} \label{s3}

We have shown that any solution of the MVP equation \eqref{e1} belongs to one of the three classes: \eqref{e7}, \eqref{e12}, and \eqref{e13}. The general solution \eqref{e7} is determined by the general solution of the sine-Gordon equation and one extra arbitrary function, while the special solutions \eqref{e12} and \eqref{e13} are given explicitly. In this sense, the MVP equation \eqref{e1} is an avatar of the sine-Gordon equation \eqref{e5}. Let us add that the MVP equation appeared also in \cite{SS} (see eq. (16) there) as an auxiliary equation, in the transformation way from the cubic Rabelo equation (a.k.a. the short pulse equation) to the sine-Gordon equation.

\end{document}